%% file: draft_1005.tex
\documentclass[amsmath,amssymb,twocolumn,prd,floatfix,showpacs, nofootinbib]{revtex4-1}
\usepackage{amsmath} 
\usepackage{graphicx} 
\usepackage{subfigure}
\usepackage{epstopdf} 
\usepackage{color} 
\usepackage{multirow}
\usepackage{setspace} 
\usepackage{overpic} 
\usepackage{amssymb}
\usepackage[normalem]{ulem}
\usepackage{natbib}
\usepackage[bookmarksnumbered, pdfstartview=FitH,colorlinks,urlcolor=blue, citecolor=blue,linkcolor=blue] {hyperref}
\usepackage{lineno}
\usepackage{bm}
\usepackage{rotating}
\usepackage{xcolor}
\usepackage{makecell}
\usepackage{mathtext}
\usepackage{mathrsfs}
\usepackage[utf8]{inputenc}
\usepackage[T1]{fontenc} 
\usepackage{tikz-feynman}  
\usetikzlibrary{arrows.meta}

\usepackage{adjustbox}

\let\oldequation\equation
\let\oldendequation\endequation
\renewenvironment{equation}
 {\linenomathNonumbers\oldequation}
 {\oldendequation\endlinenomath}

\newcommand{\BESIIIorcid}[1]{\href{https://orcid.org/#1}{\hspace*{0.1em}\raisebox{-0.45ex}{\includegraphics[width=1em]{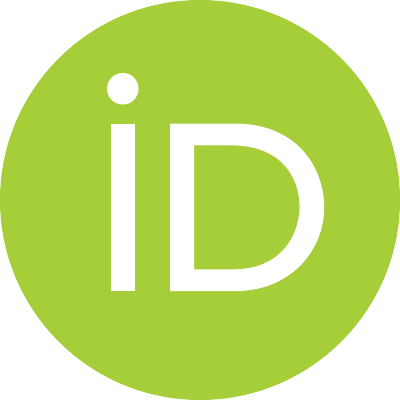}}}}

 \hypersetup{colorlinks = true,
            linkcolor = blue,
            urlcolor = blue,
            citecolor = blue,
            breaklinks=true,
            pdfstartview=Fit}

\begin{document}

\title{\boldmath First observation of the electromagnetic Dalitz decay $\psi(3686) \rightarrow \mu^+ \mu^- \eta^\prime$}
\author{
\begin{center}
\input{authorlist_2026-02-07}
\end{center}
\vspace{0.4cm}
}


\begin{abstract}
Utilizing $(2712.4 {\pm} 14.3)\times10^{6}~\psi(3686)$ events collected by the BESIII detector at the symmetric $e^+ e^-$ collider BEPCII, 
we report the first observation of the electromagnetic Dalitz decay $\psi(3686) \to \mu^+ \mu^-\eta^{\prime} $ with a statistical significance of 6.1$\sigma$. 
The branching fraction is determined to be $\mathcal{B}(\psi(3686) \to \mu^+ \mu^- \eta^{\prime})=(4.1 \pm 1.0_{\rm stat.} \pm 0.4_{\rm syst.})\times 10^{-7}$.
 The ratio to the branching fraction of the radiative decay $\psi(3686) \to \gamma \eta^{\prime}$ is estimated to be $(3.3\pm0.9)\times10^{-3}$, 
which is consistent with the prediction of the vector meson dominance model within $1\sigma$. 
Furthermore, using the branching fraction of $\psi(3686) \to e^+ e^- \eta^{\prime}$ previously measured by the BESIII experiment, 
the ratio between the muon and the electron channels is evaluated to be $0.22\pm0.07$, 
which is consistent with the calculation of the vector meson dominance model within $1\sigma$, 
and no significant violation of lepton flavor universality is found.

\end{abstract}

\maketitle

\section{Introduction}
\label{int}
\vspace{-\baselineskip}
The electromagnetic (EM) Dalitz decays $V \to \ell^{+} \ell^{-} P $ 
play an important role in understanding both the internal structure of hadrons and the fundamental mechanism of interactions between photons and hadrons~\cite{Landsberg:1985fd}, 
where $V$ is a vector meson, 
$P$ is a pseudoscalar meson and $\ell$ is a lepton ($l$ = $e$, $\mu$). 
These EM Dalitz decays proceed through a two-body radiative process of $V \to \gamma^{*} P $, 
where $\gamma^{*}$ is the virtual photon that finally decays into the lepton pair. 
Assuming $V$ and $P$ to be point-like particles~\cite{Kroll:1955zu}, 
the point-like, pure quantum-electrodynamics (QED) contribution to 
the differential transition rate of EM Dalitz decays can be accurately calculated.
The $V\to P$ transition vertex is modified by strong interaction dynamics that can be described by QCD-inspired models~\cite{Achasov:1992ku,Klingl:1996by,Faessler:1999de,Terschluesen:2010ik,Ivashyn:2011hb}. 
Usually, the differential decay width of $V \to  \ell^{+} \ell^{-} P$ is normalized to that of the corresponding radiative decay $V \to  \gamma P$, and the general form is represented by the formula
\begin{equation}
\label{equ:dGVPll}
\begin{aligned} 
  \frac{\mathrm{d} \Gamma\left( V \rightarrow  \ell^{+} \ell^{-} P  \right)}{\mathrm{d} q^{2} \, \Gamma(V \rightarrow  \gamma P )} 
  &= \left | F_{VP}(q^{2}) \right|^{2} \times {\rm QED}(q^{2}).
\end{aligned}
\end{equation}
Here, $q^2$ is the squared four-momentum transfer, e.g., the squared lepton pair invariant mass ($q^2= M^2_{\ell^+\ell^-}$), 
${\rm QED}(q^{2})$ represents the differential decay width of $V \to \ell^{+} \ell^{-} P$ from the QED calculation~\cite{Kroll:1955zu}, 
and $F_{VP}(q^{2})$ is the normalized transition form factor (TFF) with the normalization of $F_{VP}(0)=1$. 
In the vector meson dominance model (VMD), the hadronic EM current is proportional to vector meson fields~\cite{GellMann:1961tg, Bauer:1977iq}. 
Consequently, the VMD predicts that the TFF exhibits a rising behavior as a function of the dilepton invariant mass. 
In the single pole approximation, the TFF can be parameterized with the formula
\begin{equation}
\label{equ:formfactor}
F_{V\!P}(q^{2}) = \frac{1}{1 - q^{2}/\Lambda^{2}},
\end{equation}
where $\Lambda$ is an effective pole mass that takes into account the overall effects from all possible resonance poles and scattering terms in the time-like kinematic region.
The EM Dalitz decays have been extensively studied in the light-quark and charm sectors by 
SND~\cite{Achasov:2000ne}, CMD-2~\cite{Akhmetshin:2005vy}, NA60~\cite{Arnaldi:2009aa}, KLOE-2~\cite{Babusci:2014ldz}, and BESIII ~\cite{Fu:2011olo, Ablikim:2014nro, BAM:266, BAM:241}, 
with proceses of $\omega \to \ell^+ \ell^- \pi^0$, 
$\phi \to e^+ e^- \pi^0$, 
$\phi \to e^+ e^- \eta$, 
and $J/\psi \to e^{+} e^{-} \eta^{(\prime)}$. 
Most experimental results are consistent with the VMD expectations. 
However, NA60 observed a significant deviation in their measurement of the TFF for $\omega \to \mu^+\mu^- \pi^0$~\cite{Arnaldi:2009aa}, 
which shows a rapid increase close to the kinematic cutoff, 
exceeding the VMD prediction by nearly a factor of 10~\cite{ref:17}. 
Furthermore, in the study of the TFF in $J/\psi \to  e^+e^- \pi^0$~\cite{Ablikim:2014nro}, 
the $\rm BESIII$ Collaboration observed a prominent peaking structure in the $M_{e^+e^-}$ spectrum near the $\rho/\omega$ resonance mass. 
Accounting for this structure, the measured branching fraction (BF) is approximately twice the VMD prediction~\cite{ref:20, ref:21, ref:22}.

In this paper, the first experimental study of the EM Dalitz decay $\psi(3686) \to \mu^+ \mu^-\eta^{\prime}$ is reported. 
The analysis is based on a data sample of $(2712.4\pm14.3)\times10^6$ $\psi(3686)$ events~\cite{sys:njpsi} 
collected with the BESIII detector at a center-of-mass energy of $\sqrt{s}=3.686$ GeV. 
This result provides a valuable probe of the QCD phenomena 
and plays an important role in understanding the interplay between perturbative and nonperturbative effects.

\section{BESIII DETECTOR and MONTE CARLO SIMULATION}
\label{sec:BES}

The BESIII detector~\cite{Ablikim:2009aa} records symmetric $e^+e^-$ collisions 
provided by the BEPCII storage ring~\cite{Yu:IPAC2016-TUYA01}
in the center-of-mass energy range from 1.84 to 4.95~GeV,
with a peak luminosity of $1.1 \times 10^{33}\;\text{cm}^{-2}\text{s}^{-1}$ 
achieved at $\sqrt{s} = 3.773\;\text{GeV}$. 
BESIII has collected large data samples in this energy region~\cite{Ablikim:2019hff, EcmsMea, EventFilter}. 
The cylindrical core of the BESIII detector covers 93\% of the full solid angle and consists of a helium-based
 multilayer drift chamber~(MDC), a time-of-flight
system~(TOF), a CsI(Tl) electromagnetic calorimeter~(EMC), 
which are all enclosed in a superconducting solenoidal magnet
providing a 1.0~T magnetic field. 
The solenoid is surrounded by the muon counter (MUC), which is an octagonal flux-return yoke made of steel, interleaved with resistive-plate-counter muon-identification modules.
The charged-particle momentum resolution at $1~{\rm GeV}/c$ is
$0.5\%$, and the ${\rm d}E/{\rm d}x$ resolution is $6\%$ for electrons
from Bhabha scattering. 
The EMC measures photon energies with a
resolution of $2.5\%$ ($5\%$) at $1$~GeV in the barrel (end cap)
region. 
The time resolution in the plastic scintillator TOF barrel region is 68~ps, while
that in the end cap region was 110~ps. 
The end cap TOF
system was upgraded in 2015 using multigap resistive plate chamber
technology, providing a time resolution of
60~ps,
which benefits $\sim$83\% of the data used in this analysis~\cite{etof}. 
The MUC is divided into eight and four segments in the barrel and end-cap regions, respectively, 
with a total of 9 layers (5 in the barrel and 4 in the end caps). 
The total thickness of the absorber material (iron) is approximately 45 cm, 
including the coil and the yoke. 
The typical misidentification rate for pions misidentified as muons is around 1-2\%, while the detection efficiency for muons with momenta above 0.4~GeV/$c$ is better than 95\%.

Monte Carlo (MC) simulated data samples produced with a {\sc geant4}-based~\cite{geant4} software package, which includes the geometric description of the BESIII detector and the
detector response, are used to determine detection efficiencies and to estimate backgrounds. 
The simulation models the beam energy spread and initial state radiation (ISR) in the $e^+e^-$ annihilation with the generator {\sc kkmc}~\cite{ref:kkmc}.
The inclusive MC sample includes the production of the $\psi(3686)$ resonance, the ISR production of $J/\psi$, and the continuum processes incorporated in {\sc kkmc}~\cite{ref:kkmc}.
All particle decays are modeled with {\sc evtgen}~\cite{ref:evtgen} using BFs taken from the Particle Data Group (PDG)~\cite{pdg} when available, or otherwise estimated with {\sc lundcharm}~\cite{ref:lundcharm}. 
Final state radiation from charged final state particles is incorporated using the {\sc photos} package~\cite{photos2}.

The signal MC sample is utilized to optimize event selection criteria and to determine the detection efficiency. 
The $\psi(3686) \to \mu^+ \mu^- \eta^{\prime}$ decay is generated using a TFF parameterized by Eq.~(\ref{equ:formfactor}) with $\Lambda = 3.773$ GeV$/c^2$, after considering the polarization of $\psi(3686)$~\cite{ref:20, ref:21, ref:22}. 
The $\eta^\prime\to \pi^+ \pi^- \eta$ decay is modeled based on the Dalitz plot analysis of Ref.~\cite{kangxiaolin}, 
and the $\eta \to \gamma \gamma$ decay is generated uniformly in phase space.

\section{EVENT SELECTIONS}
\label{sec:selection}

The signal process targeted in this analysis is $\psi(3686)\rightarrow\mu^+\mu^-\eta^\prime$ with $\eta^\prime\rightarrow\pi^+\pi^-\eta$ and $\eta\rightarrow\gamma\gamma$.
Charged tracks are required to be within a polar angle ($\theta$) range of $|\rm{cos\theta}|<0.93$ with respect to the $z$-axis of the MDC detector, 
and to pass within $\pm 10~{\rm cm}$ of the interaction point (IP) along the $z$-direction and 
within $1~{\rm cm}$ in the plane perpendicular to the $z$-axis. 
Each event is required to contain four charged tracks with zero net charge.

Photon candidates are identified using isolated showers in the EMC. 
The deposited energy of each shower must be more than 25~MeV in the barrel region ($|\cos \theta|< 0.80$) and more than 50~MeV in the end cap regions ($0.86 <|\cos \theta|< 0.92$). 
To exclude showers that originate from charged tracks, the opening angle between the EMC shower and the closest charged track, as measured from the IP, must be greater than 10 degrees. 
To suppress electronic noise and showers unrelated to the event, the difference between the EMC time and the start time of the event is required to be within [0, 700]\,ns. 
Events with at least two photon candidates are retained for further analysis.

\begin{figure}[htbp]
    \centering
    \includegraphics[width=0.9\columnwidth]{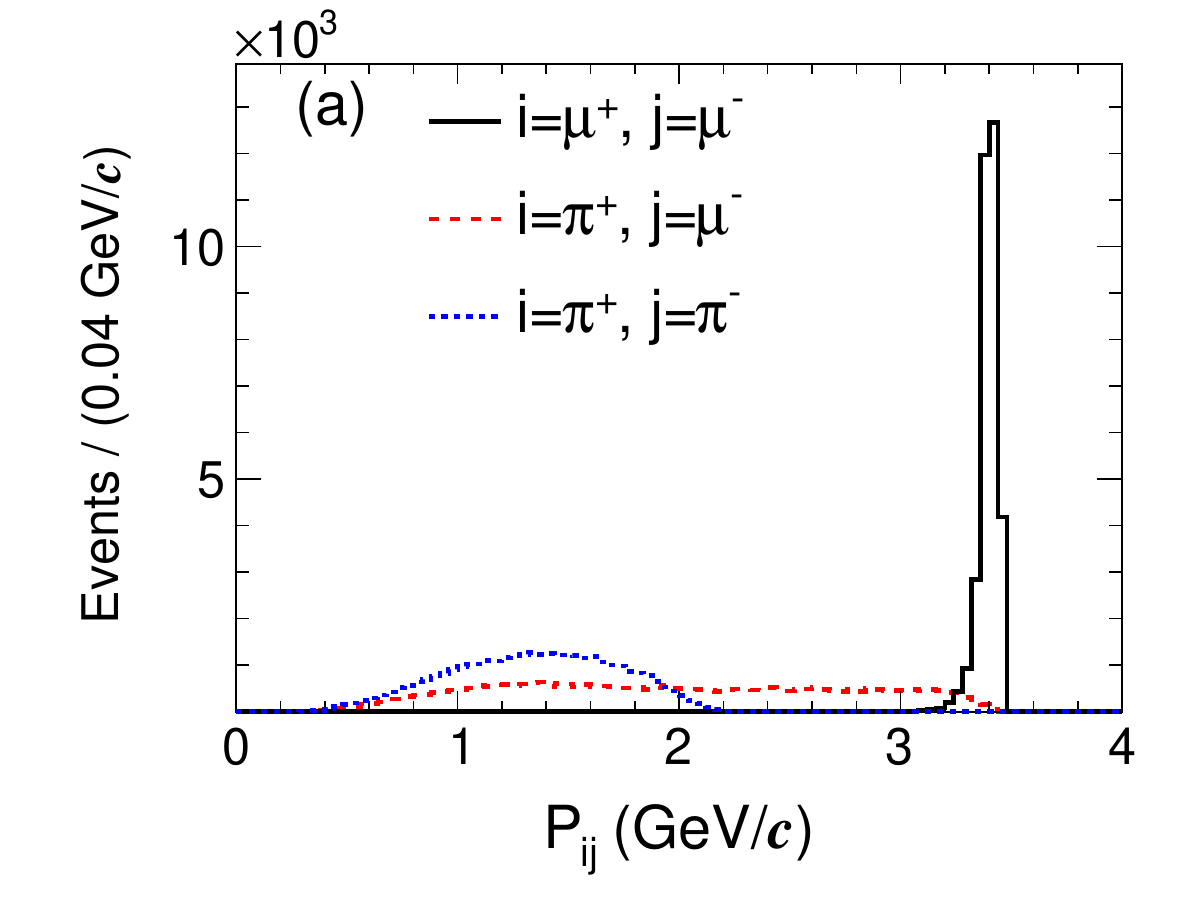}
    \vspace{0.2cm}
    \includegraphics[width=0.9\columnwidth]{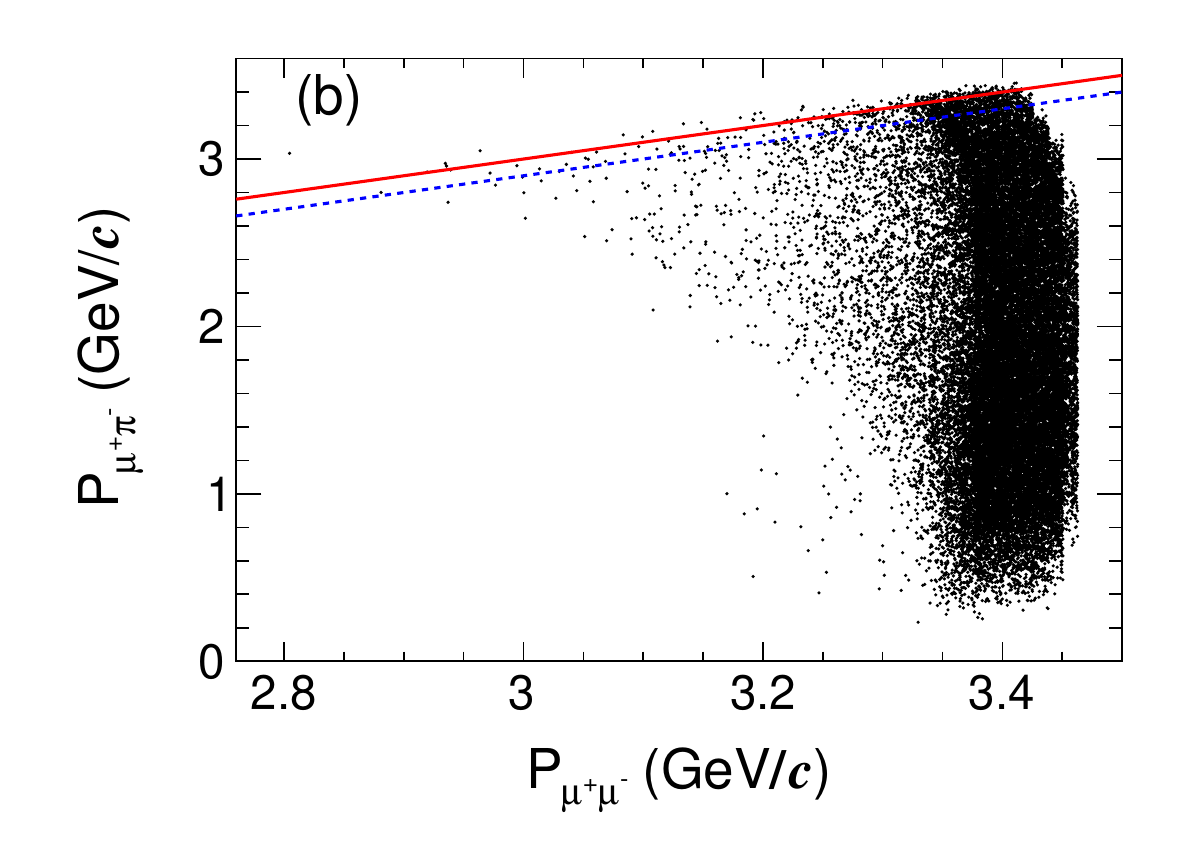}
    \caption{(a) Distributions of $P_{ij}$ corresponding to the three combinations calculated at the generator level from the signal MC sample. 
The black solid line corresponds to the $\mu^+ \mu^-$ combination, 
the red dashed line to the $\pi^+ \mu^-$ combination, 
and the blue dotted line to the $\pi^+ \pi^-$ combination. 
(b) Distribution of $P_{\mu^+ \pi^-}$ versus $P_{\mu^+ \mu^-}$. 
The red solid line represents the case where $P_{\mu^+ \pi^-}$ equals $P_{\mu^+ \mu^-}$, and the blue dashed line denotes $P_{\mu^+\mu^-} = P_{\mu^+\pi^-} +$ $0.1~{\rm GeV/}c$ case.
    }
    \label{fig:pij_1}
\end{figure}

Due to the similar masses of muons and pions, 
the particle identification (PID) information based on combined measurements of ionization energy loss ($\mathrm{d}E/\mathrm{d}x$) in the MDC 
and flight time from the TOF system is insufficient to reliably distinguish between them.
 To improve the separation, considering the different momentum distributions of muons and pions in the signal, a variable $P_{ij}$ is introduced and 
defined as $P_{ij}=|\vec{p}_i|+|\vec{p}_j|+|\vec{p}_i+\vec{p}_j|$, 
where $\vec{p}$ represents the three-momentum of the particle, $i$ refers to a positive charged track ($\mu^{+}$ or $\pi^{+}$), 
and $j$ refers to a negative charged track ($\mu^{-}$ or $\pi^{-}$). 
The distributions of $P_{ij}$ found in simulated signal events are shown in Fig.~\ref{fig:pij_1}, 
where a clear separation between the muon and pion samples can be observed. 
Among all possible pairings, $P_{ij}^{\mathrm{max}}$ denotes the maximum value of $P_{ij}$, 
while $P_{ij}^{\mathrm{others}}$ refers to the $P_{ij}$ values of the remaining pairings. 
For events with $P_{ij}^{\mathrm{max}} > P_{ij}^{\mathrm{others}} +$ $0.1~{\rm GeV/}c$, 
the $P_{ij}^{\mathrm{max}}$ corresponds to the pair of $\mu^+$ and $\mu^-$.

A kinematic fit is performed to suppress backgrounds and to improve the resolution. 
The five-constraint (5C) fit is performed under the $\psi(3686)\to\mu^+\mu^-\pi^+\pi^-\eta~(\eta\to\gamma\gamma)$ hypothesis 
by constraining the total four-momentum of the final state particles to that of the colliding beams 
and constraining the invariant mass of the photon pair ($M_{\gamma \gamma}$) to the nominal $\eta$ mass~\cite{pdg}. 
In the fit, the $\mu^+$, $\mu^-$, $\pi^+$, and $\pi^-$ particles are assigned 
according to the maximum-$P_{ij}$ selection. 
For events that do not satisfy the $P_{ij}^{\mathrm{max}} > P_{ij}^{\mathrm{others}} +$ $0.1~{\rm GeV/}c$ condition, 
the assignment of $\mu^+ \mu^- \pi^+ \pi^-$ is determined by selecting the minimum value of $\chi^{2}_{\text{tot}}$, 
where $\chi^{2}_{\text{tot}} = \chi^{2}_{5\rm C} + \chi^{2}_{\text{PID}} + \chi^{2}_{\text{vtx}}$. 
Here, the $\chi^{2}_{5\rm C}$ and the $\chi^{2}_{\text{vtx}}$ represent the goodness of the 5C kinematic fit and the vertex fit, respectively. 
The vertex fit constrains the four charged tracks ($\mu^+\mu^-\pi^+\pi^-$) to originate from a common vertex. 
The $\chi^{2}_{\text{PID}}$ denotes the goodness of the PID using TOF, d$E$/d$x$ and EMC information for the four possible assignments (assuming pion or muon for each good charged track).
 For events with more than two photons, the $\chi_{\rm 5C}^2$ is calculated for all possible photon combinations. 
The two photons from the $\eta$ decay are chosen as the pair with the minimum $\chi_{\rm 5C}^2$. 
Furthermore, $\chi^{2}_{\rm 5C}<30$ is applied, with the criterion optimized with the figure of merit of $S/\sqrt{S+B}$, 
where $S$ and $B$ denote the numbers of signal and background events from signal and inclusive MC samples, respectively. 
Subsequently, the $\eta^\prime$ candidate is reconstructed from the selected $\eta$, $\pi^+$ and $\pi^-$ particles.

According to the analysis of the signal MC sample, 
approximately 95\% of events satisfy the $P_{ij}^{\mathrm{max}} > P_{ij}^{\mathrm{others}} +$ $0.1~{\rm GeV/}c$ condition. 
When applying this maximum-$P_{ij}$ selection, the $\mu$-$\pi$ misidentification rate is found to be less than $0.1\%$. 
In contrast, the misidentification rate rises to approximately $18.6\%$ when only the $\chi^{2}_{\rm tot}$ selection is employed.

After applying the above selection criteria, there is a large background component from the $\psi(3686) \to \pi^+ \pi^- \eta^{\prime}$, 
where $\pi^+ \pi^-$ mainly comes from $\rho$ resonance, due to misidentification between pions and muons. 
To suppress this dominant peaking background, at least one muon candidate has to satisfy further muon identification criteria.
 The number of hit layers in the MUC for $\mu^{\pm}$ is required to be at least 3; 
for tracks in the momentum range $|\vec{p}_{\mu^{\pm}}| \in [0.6, 1.1]~\text{GeV}/c$, 
the penetration depth in the MUC detector (MUC depth) is required to be greater than ($68\cdot |\vec{p}_{\mu^{\pm}}|/({\rm GeV}/c) -37.4)~\text{cm}$; 
for tracks with $|\vec{p}_{\mu^{\pm }}| > 1.1$~GeV$/c$, the MUC depth is required to be greater than 37.4 cm.

\begin{figure*}[ht]
  \includegraphics[width=1\textwidth, clip, trim = {0 0.5cm 0 0}]{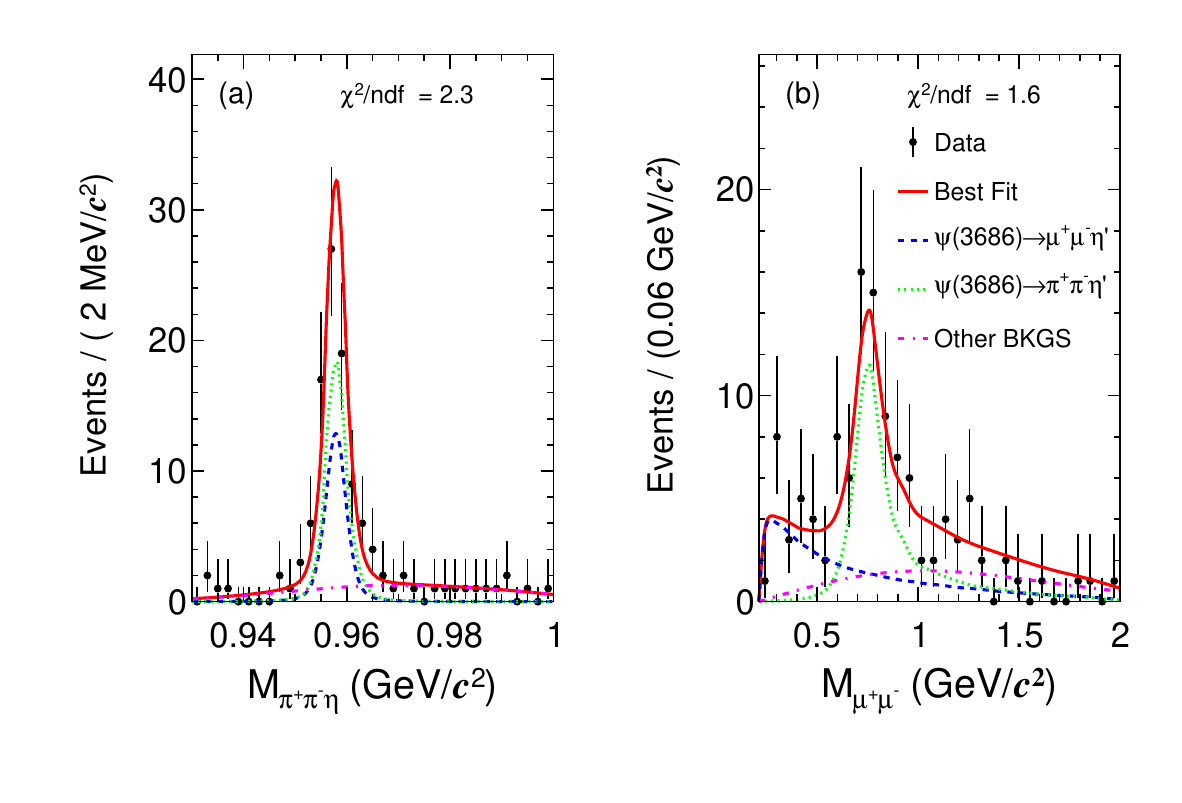}
  \caption{Projections of the 2D fit to (a) $\mathrm{M}_{\pi^+ \pi^- \eta}$ and (b) $\mathrm{M}_{\mu^+ \mu^-}$. 
The black dots with error bars are data, and the red solid line represents the best fit. 
The blue dashed line denotes the signal $\psi(3686) \to \mu^+ \mu^- \eta^{\prime}$ components, 
the green dotted line denotes the main peaking background from $\psi(3686) \to \pi^+ \pi^- \eta^{\prime}$, 
and the magenta dash-dotted line denotes the other backgrounds. 
}
  \label{fig:eeRec_fit_data}
\end{figure*}

\section{BF measurement}
The signal yield of $\psi(3686)\rightarrow\mu^+\mu^-\eta^{\prime}$ is 
extracted from a two-dimensional (2D) unbinned maximum likelihood fit to the $M_{\pi^+ \pi^- \eta}$ and $M_{\mu^{+}\mu^{-}}$ distributions. 
The total probability density function consists of a signal component and various background contributions. 
The signal component is modeled using the shape obtained from MC simulations. 
The background contributions are modeled with two components: 
(i) the main peaking background $\psi(3686) \to \pi^{+} \pi^{-} \eta^{\prime}$ 
due to pions misidentified as muons, 
is estimated using the results from a partial wave analysis of $\psi(3686)$ decays to $\pi^+\pi^-\eta^\prime$~\cite{pipietap3686}; 
(ii) other backgrounds are primarily due to misidentification of pions as muons in non-$\eta^{\prime}$ production processes, 
and are described using the inclusive MC sample after excluding the main peaking background.

Figure~\ref{fig:eeRec_fit_data} shows the projections of the 2D fitting results.
The $\chi^2/{\rm ndf}$ is also displayed on each plot to demonstrate the goodness of fit, where the ndf means the number of degree of freedom. 
The number of signal $\psi(3686)\rightarrow\mu^+\mu^-\eta^{\prime}$ events is obtained to be $N_{\mathrm{sig}}=35 \pm 8$. 
The statistical significance of the signal is evaluated to be $6.1\sigma$ with the formula of $\mathcal{S}=\sqrt{-2\ln(\mathcal{L}_0/\mathcal{L}_{\rm max})}$,
where $\mathcal{L}_{\rm max}$ is the maximum likelihood from the nominal fit,
and $\mathcal{L}_0$ is the likelihood value when $N_{\mathrm{sig}}$ is fixed at 0.

The BF of $\psi(3686)\rightarrow\mu^+\mu^-\eta^{\prime}$ is calculated by
\begin{equation}
\label{equ:branch}
   \mathcal{B}(\psi(3686) \to \mu^+ \mu^- \eta^{\prime}) = \dfrac{ N_{\rm sig} }{ N_{\psi(3686)} \cdot \varepsilon_{\rm sig}  \cdot \mathcal{B}_{1} \cdot \mathcal{B}_{2}},
\end{equation}
where $N_{\psi(3686)}$ is the total number of $\psi(3686)$ events of data, 
$\varepsilon_{\rm sig}=(18.6\pm0.03)$\% is the detection efficiency obtained from the signal MC sample, 
and $\mathcal{B}_{1}=(42.5 \pm 0.5)\%$ and $\mathcal{B}_{2}=(39.36 \pm 0.18)\%$
 are the BFs of $\eta^{\prime} \to \pi^+ \pi^- \eta$ and $\eta \to \gamma \gamma$, respectively, 
quoted from the PDG~\cite{pdg}. 
Using the measured signal yield, the BF of $\psi(3686) \to \mu^+ \mu^- \eta^{\prime}$ is 
determined to be $ (4.1 \pm 1.0_{\rm stat.} \pm 0.4_{\rm syst.}) \times 10^{-7}$. 
The systematic uncertainties are discussed in Section~\ref{system}. 
The ratio of this BF to that of the radiative decay is determined using the world average $\mathcal{B}(\psi(3686) \to \gamma\eta^\prime) = (1.24 \pm 0.04) \times 10^{-4}$~\cite{pdg}, yielding
\begin{equation}
\label{equ:branch1}
R_{1} \equiv \frac{\mathcal{B}(\psi(3686) \to \mu^+\mu^-\eta^\prime)}{\mathcal{B}(\psi(3686) \to \gamma\eta^\prime)} = (3.3 \pm 0.9) \times 10^{-3}.
\end{equation}
It is consistent with the VMD prediction of $2.96 \times 10^{-3}$ based on the single pole approximation. 
In addition, the ratio of the BFs between the muonic and electronic channels 
is evaluated using the previous BESIII measurement $\mathcal{B}(\psi(3686) \to e^+e^-\eta^\prime) = (1.90 \pm 0.25_{\rm stat.} \pm 0.11_{\rm syst.}) \times 10^{-6}$~\cite{eeetap3686}, giving
\begin{equation}
\label{equ:branch2}
R_{2} \equiv \frac{\mathcal{B}(\psi(3686) \to \mu^+\mu^-\eta^\prime)}{\mathcal{B}(\psi(3686) \to e^+e^-\eta^\prime)} = 0.22 \pm 0.07.
\end{equation}
This result is consistent with the theoretical prediction of $0.26$ based on the VMD model. 
The uncertainties in the $R_{1}$ and $R_{2}$ results are dominated by the statistical uncertainties from the $\psi(3686) \to \mu^{+}\mu^{-} \eta'$ channel.

\section{Systematic Uncertainties}
\label{system}
The sources of systematic uncertainties include the MDC tracking, the photon detection, 
the muon identification, the kinematic fit, the signal model, the shapes of peaking and other backgrounds, 
the BFs quoted from the PDG and the total number of $\psi(3686)$ events. 
Their effects on the BF measurement are presented below.

\begin{itemize}

  \item 

   The MDC tracking efficiency of pion has been studied with the control sample of $J/\psi \to \rho \pi$~\cite{sys:helix1}. 
The efficiency difference of pion between data and MC is 1.0\% per track. 
The systematic uncertainty from the tracking of $\mu^{\pm}$ is determined to be 1.0$\%$ per track 
in the analysis of $\eta/\eta^{\prime} \to l^+ l^- \mu^+ \mu^-$~\cite{sys:helix}.

    \item 
    
    The systematic uncertainty from the photon detection has been studied in the processes $J/\psi\rightarrow\pi^+\pi^-\pi^0$~\cite{sys:Ref2}, 
and determined to be 0.5\% per photon in the EMC barrel and 1.5\% per photon in the EMC end caps. 
After weighted according to the polar angle distributions of two photons from signal MC sample, 
the systematic uncertainty from the photon detection is estimated to be 1.0\% per photon.  

\item 

    The systematic uncertainty from the muon identification in the MUC has been studied by analyzing a control sample of $e^+ e^- \to \gamma \mu^+ \mu^-$. 
The data-MC difference in the identification efficiency, 4.4\%, is taken as the systematic uncertainty.
    
   \item 
   
   The systematic uncertainty due to the kinematic fit is estimated 
by comparing the detection efficiencies with and without applying the track helix parameters following the method described in Ref.~\cite{sys:Ref1}, which is 1.2\%.

    \item 
    The systematic uncertainty from the signal shape is evaluated by generating an alternative MC sample, 
in which the pole mass $\Lambda$ value is changed from the nominal value of 3.773 GeV$/c^2$ to 4.25 GeV$/c^2$. 
The difference in detection efficiencies with the nominal signal MC sample, 0.6\%, is assigned to be the systematic uncertainty.

    \item 

    The systematic uncertainty from the background shape is estimated by replacing the inclusive MC shape with a second-order Chebyshev polynomial function, 
and by substituting the $\psi(3686) \to \pi^+ \pi^- \eta^{\prime}$ shape generated by the partial wave model~\cite{pipietap3686} 
with the processes $\psi(3686) \to \pi^+ \pi^- \eta^\prime$ and $\psi(3686) \to \rho(770) \eta^\prime$ simulated uniformly in phase space. 
The resulting difference in the measured BF, 5.3\%, is associated as the systematic uncertainty.

 \item 

    The BFs of $\eta^\prime\rightarrow\pi^+\pi^-\eta$ and $\eta\rightarrow\gamma\gamma$ are quoted from the PDG~\cite{pdg}. 
The respective uncertainties of 1.2\% and 0.5\% are taken as the systematic uncertainties.

    \item  
    
    The total number of $\psi(3686)$ events in data, 
    (2712.4 $\pm$ 14.3) $\times$ 10$^{6}$~\cite{sys:njpsi}, is determined from an analysis of inclusive $\psi(3686)$ hadronic events, and its uncertainty is 0.5\%.

\end{itemize}

\begin{table}[htbp]
    \centering
    \caption{Relative systematic uncertainties in the BF measurement (in percent).}
    \renewcommand{\arraystretch}{1.2}
    \adjustbox{max width=\textwidth}{%
        \begin{tabular}{lcc}
            \hline\hline
            Source & ~~~~~Uncertainty (\%) \\
            \hline
            MDC tracking           & $4.0$ \\
            Photon detection       & $2.0$ \\
            Muon identification    & $4.4$ \\
            Kinematic fit & $1.2$  \\
            Signal shape           & $0.6$ \\
            Background shape       & $5.3$ \\
            $\mathcal{B}(\eta^\prime\rightarrow\pi^+\pi^-\eta)$ & $1.2$ \\
            $\mathcal{B}(\eta\rightarrow\gamma\gamma)$ & $0.5$ \\
            Number of $\psi(3686)$ events & $0.5$ \\
            \hline
            Total & $8.5$ \\
            \hline\hline
        \end{tabular}%
    }
    \label{eemumusys}
\end{table}

Table~\ref{eemumusys} summarizes the above systematic uncertainties on the BF measurement. 
Assuming that all of sources are independent, the total systematic uncertainty is obtained to be 8.5\% by adding each individual contribution in quadrature.

\section{Summary}

With a sample of (2712.4 $\pm$ 14.3) $\times$ 10$^6$ $\psi(3686)$ events~\cite{sys:njpsi} collected with the BESIII detector, 
the $\psi(3686)\to\mu^+\mu^-\eta^\prime$ decay is observed via $\eta^\prime\to\pi^+\pi^-\eta$ and $\eta \to \gamma \gamma$, 
with a statistical significance of $6.1\sigma$. 
Under the assumption that the $\psi(3686)\to\eta^\prime$ TFF 
can be modeled according to Eq.~(\ref{equ:formfactor}) with a pole mass of $\Lambda =3.773~{\rm GeV}/c^2$, 
the BF is measured to be $ (4.1 \pm 1.0_{\rm stat.} \pm 0.4_{\rm syst.}) \times 10^{-7}$. 
The ratio of the Dalitz decay to the radiative decay, 
$R_1 = \mathcal{B}(\psi(3686) \to \mu^+\mu^-\eta^\prime) / \mathcal{B}(\psi(3686) \to \gamma\eta^\prime)$, 
is estimated to be $(3.3 \pm 0.9) \times 10^{-3}$. 
The ratio between the muonic and electronic decay channels, 
$R_2 = \mathcal{B}(\psi(3686) \to \mu^+\mu^-\eta^\prime) / \mathcal{B}(\psi(3686) \to e^+e^-\eta^\prime)$, 
is evaluated to be $0.22 \pm 0.07$. 
The statistical uncertainty from the $\psi(3686) \to \mu^{+}\mu^{-} \eta'$ decay
 dominates the uncertainties of the $R_{1}$ and $R_{2}$ results. 
Both ratios are consistent with the VMD predictions within $1\sigma$, 
and no significant violation of lepton flavor universality is observed. 
Although the current statistics do not allow for a differential measurement of the TFF, 
this analysis provides a foundation for future studies. 
At next-generation facilities like the Super $\tau$-Charm Facility~\cite{Achasov_2023}, 
where orders of magnitude larger $\psi(3686)$ data samples are anticipated, 
a precise determination of the TFF will be feasible, 
offering deeper insights into the EM structure of the $\psi(3686) \to \eta^\prime$ transition and enabling more stringent tests of QCD models.

\section{Acknowledgement}
The BESIII Collaboration thanks the staff of BEPCII  (https://cstr.cn/31109.02.BEPC) and the IHEP computing center for their strong support.
This work is supported in part by National Key R\&D Program of China under Contracts Nos. 2025YFA1613900, 2023YFA1606000, 2023YFA1606704; 
Outstanding Youth Foundation of Henan Province No. 252300421219; 
National Natural Science Foundation of China (NSFC) under Contracts Nos. 
12305102, 12275067, 12275068, 12105077, 
11635010, 11935015, 11935016, 11935018, 12025502, 12035009, 12035013, 12061131003, 12192260, 12192261, 12192262, 12192263, 12192264, 12192265, 12221005, 12225509, 12235017, 12342502, 12361141819, 12535005; 
the Chinese Academy of Sciences (CAS) Large-Scale Scientific Facility Program; 
the Strategic Priority Research Program of Chinese Academy of Sciences under Contract No. XDA0480600; 
CAS under Contract No. YSBR-101; 100 Talents Program of CAS; 
The Institute of Nuclear and Particle Physics (INPAC) and Shanghai Key Laboratory for Particle Physics and Cosmology; 
Science and Technology R$\&$D Program Joint Fund Project of Henan Province No.225200810030, Science and Technology Innovation Leading Talent Support Program of Henan Province No. 254000510039; 
Natural Science Foundation of Henan Province Nos. 252300421491, 242300420250; 
Agencia Nacional de Investigacion y Desarrollo de Chile (ANID), Chile under Contract No. ANID CCTVal CIA250027; 
ERC under Contract No. 758462; 
German Research Foundation DFG under Contract No. FOR5327; 
Istituto Nazionale di Fisica Nucleare, Italy; Knut and Alice Wallenberg Foundation under Contracts Nos. 2021.0174, 2021.0299, 2023.0315; Ministry of Development of Turkey under Contract No. DPT2006K-120470; National Research Foundation of Korea under Contract No. NRF-2022R1A2C1092335; National Science and Technology fund of Mongolia; Polish National Science Centre under Contract No. 2024/53/B/ST2/00975; STFC (United Kingdom); Swedish Research Council under Contract No. 2019.04595; U. S. Department of Energy under Contract No. DE-FG02-05ER41374.

\bibliographystyle{apsrev4-2}

\end{document}

%% file: authorlist_2026-02-07.tex
M.~Ablikim$^{1}$\BESIIIorcid{0000-0002-3935-619X},
M.~N.~Achasov$^{4,c}$\BESIIIorcid{0000-0002-9400-8622},
P.~Adlarson$^{83}$\BESIIIorcid{0000-0001-6280-3851},
X.~C.~Ai$^{89}$\BESIIIorcid{0000-0003-3856-2415},
C.~S.~Akondi$^{31A,31B}$\BESIIIorcid{0000-0001-6303-5217},
R.~Aliberti$^{39}$\BESIIIorcid{0000-0003-3500-4012},
A.~Amoroso$^{82A,82C}$\BESIIIorcid{0000-0002-3095-8610},
Q.~An$^{78,65,\dagger}$,
Y.~H.~An$^{89}$\BESIIIorcid{0009-0008-3419-0849},
M.~S.~Anderson$^{39}$\BESIIIorcid{0009-0008-1550-2632}
% ...
\\
\vspace{0.2cm}
(BESIII Collaboration)\\
\vspace{0.2cm} {\it
$^{1}$ Institute of High Energy Physics, Beijing 100049, People's Republic of China\\
% ...
$^{4}$ Budker Institute of Nuclear Physics SB RAS (BINP), Novosibirsk 630090, Russia\\
% ...
$^{31}$ INFN Sezione di Ferrara, (A)INFN Sezione di Ferrara, I-44122, Ferrara, Italy; (B)University of Ferrara, I-44122, Ferrara, Italy\\
% ...
$^{39}$ Johannes Gutenberg University of Mainz, Johann-Joachim-Becher-Weg 45, D-55099 Mainz, Germany\\
% ...
$^{65}$ State Key Laboratory of Particle Detection and Electronics, Beijing 100049, Hefei 230026, People's Republic of China\\
% ...
$^{78}$ University of Science and Technology of China, Hefei 230026, People's Republic of China\\
% ...
$^{82}$ University of Turin and INFN, (A)University of Turin, I-10125, Turin, Italy; (B)University of Eastern Piedmont, I-15121, Alessandria, Italy; (C)INFN, I-10125, Turin, Italy\\
$^{83}$ Uppsala University, Box 516, SE-75120 Uppsala, Sweden\\
% ...
$^{89}$ Zhengzhou University, Zhengzhou 450001, People's Republic of China\\

\vspace{0.2cm}
$^{\dagger}$ Deceased\\
% ...
$^{c}$ Also at the Novosibirsk State University, Novosibirsk, 630090, Russia\\
% ...

}